\documentclass[preprint,showpacs,preprintnumbers,amsmath,amssymb]{revtex4}
 \usepackage{graphicx}
 \usepackage{dcolumn}
 \usepackage{bm}
\begin{document}

\title{Production of exotic isotopes in complete fusion reactions with radioactive beams}
\author{V.V.~Sargsyan$^{1,2}$, A.S.~Zubov$^{1}$, G.G.~Adamian$^{1}$, N.V.~Antonenko$^1$, and S.~Heinz$^3$
}
\affiliation{$^{1}$Joint Institute for
Nuclear Research, 141980 Dubna, Russia\\
$^{2}$International Center for Advanced Studies, Yerevan State University,  0025 Yerevan, Armenia\\
$^{3}$GSI Helmholtzzentrum f\"ur Schwerionenforschung GmbH, 64291 Darmstadt, Germany
}
\date{\today}

\begin{abstract}
The  isotopic dependence of the
complete fusion (capture) cross section  is analyzed in the reactions
$^{130,132,134,136,138,140,142,144,146,148,150}$Xe+$^{48}$Ca
with stable and radioactive beams. It is shown for the first time
that the very neutron-rich nuclei $^{186-191}$W can be reached with  relatively  large cross sections
by complete fusion reactions with radioactive ion beams at incident energies
near the Coulomb barrier. A comparison between the complete fusion
and fragmentation reactions for the production of neutron-rich W
and neutron-deficient Rn  isotopes is performed.
\end{abstract}


\maketitle

\section{Introduction}
The new generation of radioactive ion beam facilities will provide high
intensity ($>10^{9}$ ions/s) exotic beams
(for example,  $^{88-94}$Kr, $^{126-132}$Sn, $^{138-144}$Xe or $^{119-132}$Cs).
One of the most interesting areas of
research with radioactive beams will be the study of the complete fusion process~\cite{Love} where
the fusion experiments with exotic beams can be performed to synthesize and study new isotopes
of existing elements.
The central issue is whether the capture and fusion cross sections will be enhanced due
to the large deformation of the neutron-rich or neutron-deficient projectile-nucleus.
However, one should
bear in mind the smaller intensity of these beams in comparison with the
intensity of stable beams.
Our aim is to find
the global trend in the production cross section of exotic nuclei as a function of
the charge (mass) number of the projectile in complete fusion reactions.
Based on this trend one can find a consensus between
the cross sections resulting from a certain beam and the intensity of this beam.

The goal of the present paper is to compare the fusion of stable $^{130,132,134,136}$Xe and
radioactive $^{138,140,142,144,146,148,150}$Xe beams with the same target, $^{48}$Ca,
in order to study the effects of the neutron excess and neutron transfer on the fusion process.
The target $^{48}$Ca is ideal for this purpose since this nucleus has the largest possible
neutron excess and the systems $^{138,140,142,144,146,148,150}$Xe+$^{48}$Ca
have  positive neutron transfer $Q$-values while all the corresponding reactions
$^{130,132,134,136}$Xe+$^{48}$Ca display negative $Q$-values.
In the present paper
we demonstrate for the first time the
possibilities for producing neutron-rich isotopes of
$^{186-191}$W in the complete fusion reactions
$^{146,148}$Xe+$^{48}$Ca with rather large cross sections.

The nucleus $^{190}$W  was the heaviest isotope which has been synthesized in ($n$,$n2p$) and
($p$,$3p$) reactions \cite{NPN}. In  these experiments the chemical extraction of $^{190}$W was possible after
long irradiation. Another method to produce the neutron-rich nuclei
is fragmentation reactions~\cite{Ben,Pod}. Cross-sections smaller than 0.4$\mu$b were measured for the isotopes
$^{190-192}$W in cold fragmentation of 950 MeV/nucleon $^{197}$Au beams on Be targets \cite{Ben}.
However, the production cross section decreases strongly with increasing neutron number.
The most neutron-rich W isotopes, up to $^{197}$W, were observed in projectile fragmentation of
$^{238}$U at 1 GeV/nucleon on Be targets at the Fragment Separator (FRS) at GSI \cite{Kurc}.
Here, cross-sections smaller than 5 nb were measured
for W isotopes with mass numbers A $\geq$ 190 where the cross-section decreases by approximately one
order of magnitude for every two neutrons more in the residual nucleus. In the present paper
we also compare the complete fusion reactions $^{146}$Xe+$^{48}$Ca with fragmentation reactions
leading both to the production of neutron-rich W isotopes. Additionally, we will compare the
complete fusion reactions $^{123}$Cs+$^{69}$Ga which lead to neutron-deficient Rn isotopes with
the respective yields from the fragmentation reactions.

\section{Model}
Because
the capture cross section is equal to the fusion cross section for the reactions $^{A}$Xe+$^{48}$Ca treated
in the present paper, the quantum diffusion approach \cite{EPJSub,EPJSub1} for the capture is applied to study
the complete fusion.
With this approach many heavy-ion capture
reactions at energies above and well below the Coulomb barrier have been
successfully described~\cite{EPJSub,EPJSub1,PRCPOP}.
Since the details of our theoretical treatment were already published in
Refs. \cite{EPJSub,EPJSub1,PRCPOP}, the model will be only shortly described.

In the quantum diffusion approach \cite{EPJSub,EPJSub1}
the collisions of  nuclei are described with
a single relevant collective variable: the relative distance between
the colliding nuclei. This approach takes into consideration the fluctuation and dissipation effects in
collisions of heavy ions which model the coupling with various channels
(for example, coupling of the relative motion with low-lying collective modes
such as dynamical quadrupole and octupole modes of the target and projectile \cite{Ayik333}).
We have to mention that many quantum-mechanical and non-Markovian effects accompanying
the passage through the Coulomb barrier are taken into consideration in our
formalism \cite{EPJSub,EPJSub1,PRCPOP}. The diffusion models, which  include the quantum statistical effects,
were also proposed in Refs. \cite{Hofman}.
The  nuclear deformation effects
are taken into account through the dependence of the nucleus-nucleus potential
on the deformations and mutual orientations of the colliding nuclei.
To calculate the nucleus-nucleus interaction potential $V(R)$,
we use the procedure presented in Ref. \cite{EPJSub1}.
For the nuclear part of the nucleus-nucleus potential, the double-folding formalism with
a Skyrme-type density-dependent effective nucleon-nucleon interaction is used \cite{Adamian96}.
The nucleon densities of the projectile and target nuclei
are specified in the form of the Woods-Saxon parameterization,
where the nuclear radius parameter is $r_0=1.15$  fm  and the
diffuseness parameter takes the values $a= 0.55$ fm  for all nuclei.
The absolute values of the quadrupole deformation parameters $\beta_2$
of deformed nuclei were taken from Refs. \cite{Ram} and \cite{MN}
  for the known and unknown nuclei, respectively.
 For the magic $^{48}$Ca and
semimagic $^{136}$Xe nuclei in the ground state, we set $\beta_2=0$ and $\beta_2=0.05$, respectively.

The capture cross section is the sum of the partial capture cross sections \cite{EPJSub,EPJSub1}
\begin{eqnarray}
\sigma_{cap}(E_{\rm c.m.})&=&\sum_{J}^{}\sigma_{\rm cap}(E_{\rm
c.m.},J)=\nonumber\\&=& \pi\lambdabar^2
\sum_{J}^{}(2J+1)\int_0^{\pi/2}d\theta_1\sin\theta_1\int_0^{\pi/2}d\theta_2\sin\theta_2 P_{\rm cap}(E_{\rm
c.m.},J,\theta_1,\theta_2),
\label{1a_eq}
\end{eqnarray}
where $\lambdabar^2=\hbar^2/(2\mu E_{\rm c.m.})$ is the reduced de Broglie wavelength,
$\mu=m_0A_1A_2/(A_1+A_2)$ is the reduced mass ($m_0$ is the nucleon mass),
and the summation is over the possible values of the angular momentum $J$
at a given bombarding energy $E_{\rm c.m.}$.
Knowing the potential of the interacting nuclei for each orientation with the angles $\theta_i (i=1,2)$, one can obtain the partial capture probability
$P_{\rm cap}$ which is defined by the probability to penetrate the potential barrier in the relative distance coordinate $R$
 at a given $J$.
The value of $P_{\rm cap}$
is obtained by integrating the propagator $G$ from the initial
state $(R_0,P_0)$ at time $t=0$ to the final state $(R,P)$ at time $t$ ($P$ is the momentum):
\begin{eqnarray}
P_{\rm cap}&=&\lim_{t\to\infty}\int_{-\infty}^{r_{\rm in}}dR\int_{-\infty}^{\infty}dP\  G(R,P,t|R_0,P_0,0)\nonumber \\
&=&\lim_{t\to\infty}\frac{1}{2} {\rm erfc}\left[\frac{-r_{\rm in}+\overline{R(t)}}
{{\sqrt{\Sigma_{RR}(t)}}}\right].
\label{1ab_eq}
\end{eqnarray}
Here, $r_{\rm in}$ is an internal turning point.
The second line in (\ref{1ab_eq}) is obtained by using the propagator
$G=\pi^{-1}|\det {\bf \Sigma}^{-1}|^{1/2}
\exp(-{\bf q}^{T}{\bf \Sigma}^{-1}{\bm q})$
(${\bf q}^{T}=[q_R,q_P]$,
$q_R(t)=R-\overline{R(t)}$, $q_P(t)=P-\overline{P(t)}$, $\overline{R(t=0)}=R_0$,
$\overline{P(t=0)}=P_0$, $\Sigma_{kk'}(t)=2\overline{q_k(t)q_{k'}(t)}$, $\Sigma_{kk'}(t=0)=0$,
$k,k'=R,P$) calculated  for
an inverted oscillator which approximates
the nucleus-nucleus potential $V$ in the variable $R$ as follows.
At given $E_{\rm c.m.}$ and $J$, the
classical action is calculated for the realistic nucleus-nucleus potential.
Then the   realistic nucleus-nucleus potential is replaced by an inverted oscillator
which has the same barrier height and classical action. So, the frequency
 $\omega(E_{\rm c.m.},J)$ of
this oscillator is set to obtain an equality of the
classical actions in the approximated and realistic potentials.
The action is calculated in the WKB approximation which is the accurate
at the sub-barrier energies.
Usually in the literature the parabolic approximation with
$E_{\rm c.m.}$-independent $\omega$ is employed
which is not accurate at the deep sub-barrier energies.
Our approximation is well justified for the
reactions and energy range considered here \cite{EPJSub,EPJSub1}.

We assume that the sub-barrier capture  mainly  depends  on the
two-neutron
transfer with positive  $Q$-value.
Our assumption is that, just before the projectile is captured by the target-nucleus
(just before the crossing of the Coulomb barrier),
the  transfer  occurs   and  leads to the
population of the first excited collective state in the recipient nucleus \cite{SSzilner}.
So, the motion to the
$N/Z$ equilibrium starts in the system before the capture
because it is energetically favorable in the dinuclear system in the vicinity of the Coulomb barrier.
For the reactions under consideration,
the average change of mass asymmetry is connected to the
two-neutron
transfer.
Since after the transfer the mass numbers, the isotopic composition and the deformation parameters
of the interacting nuclei, and, correspondingly, the height $V_b=V(R_b)$
and shape of the Coulomb barrier are changed,
one can expect an enhancement or suppression of the capture.
If  after the neutron transfer the deformations of the interacting nuclei increase (decrease),
the capture probability increases (decreases).
When the isotopic dependence of the nucleus-nucleus
potential is weak and   after the transfer the deformations of the interacting nuclei do not change,
there is no effect of the neutron transfer on the capture.
In comparison with Ref. \cite{Dasso}, we assume that the negative transfer $Q-$values
do not play  a visible  role in the capture process.
Our scenario was verified in the description of many reactions \cite{EPJSub1}.

The primary
neutron-rich products of  the complete fusion reactions $^{A}$Xe+$^{48}$Ca of interest
are excited and  transformed into the secondary products with
a smaller number of neutrons.
Since neutron emission is the dominant
deexcitation channel in the neutron-rich isotopes of interest, the production cross sections
of the secondary nuclei are the same as those of the corresponding primary nuclei.
This seems to be evident without special statistical treatment.
The calculations are performed
by employing the predicted values of the mass excesses and the neutron separation energies
$S_n(Z,N)$ for unknown nuclei   from the finite range liquid drop model \cite{MN}.

\section{Results of the calculations}
\subsection{Complete fusion reactions $^{A}$Xe+$^{48}$Ca}
To analyze the isotopic trend of the fusion  cross section, it is useful to use the so called
universal fusion function (UFF) representation \cite{GomesUFF}. The advantage of  this representation
appears clearly when one wants to compare fusion  cross sections for systems with  different
Coulomb barrier heights and positions. In the reactions where the capture and fusion cross sections
coincide, the elimination of the influence
of the nucleus-nucleus
potential on the fusion cross section with the UFF representation
allows us to  conclude about the role of deformation of the
colliding nuclei and the nucleon transfer between interacting nuclei in the capture and fusion.
\begin{figure}[tbp]
\includegraphics[scale=1]{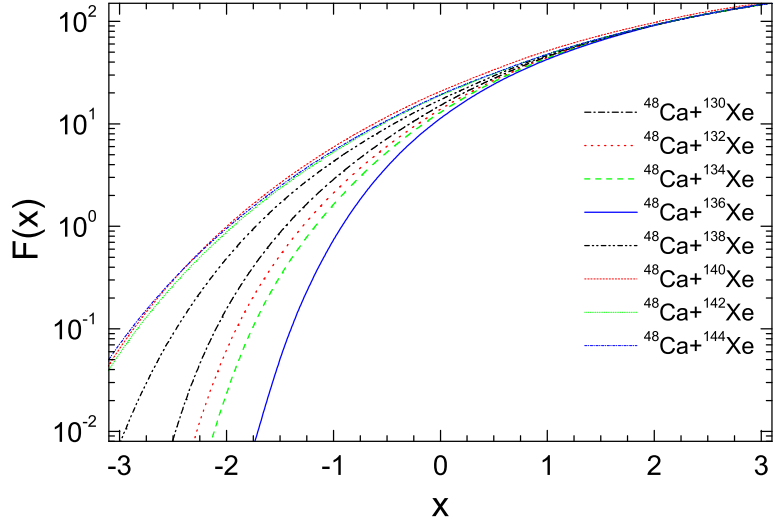}
\caption{(Color online) Calculated dependencies of
$F(x)=\frac{2E_{\rm c.m.}}{\hbar\omega_bR_b^2}\sigma$ on $x=\frac{E_{\rm c.m.}-V_b}{\hbar\omega_b}$
for the indicated reactions.}
\label{1_fig}
\end{figure}
\begin{figure}[tbp]
\includegraphics[scale=1]{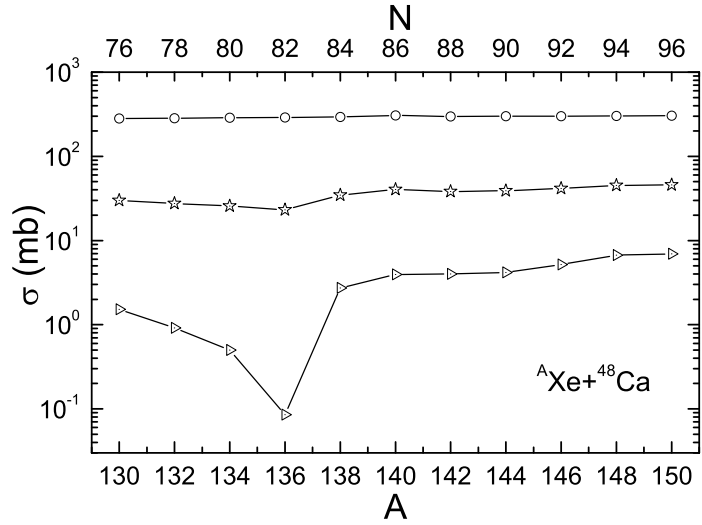}
\caption{Calculated dependence of fusion cross section
$\sigma$ on $A$
for the  reactions $^{A}$Xe+$^{48}$Ca at fixed bombarding energies
$E_{\rm c.m.}=V_b-5$ MeV (triangles), $V_b$ (stars), $V_b+10$ MeV (circles).}
\label{2_fig}
\end{figure}
In Ref. \cite{GomesUFF} the reduction procedure consists of the following transformations:
$$E_{\rm c.m.} \rightarrow x= \frac{E_{\rm c.m.}-V_b}{\hbar \omega_b},
\qquad \sigma \rightarrow F(x)=\frac{2 E_{\rm c.m.}}{\hbar \omega_b R_b^{2}}\sigma.$$
The frequency $\omega_b =\sqrt{|V^{''}(R_b)|/\mu}$
is related with the second derivative $V^{''}(R_b)$ of the nucleus-nucleus potential $V(R)$
at the barrier radius $R_b$ and the reduced mass parameter $\mu$.
With these replacements one can compare the cross sections for different reactions.
%
%
\begin{figure}[tbp]
\includegraphics[scale=1]{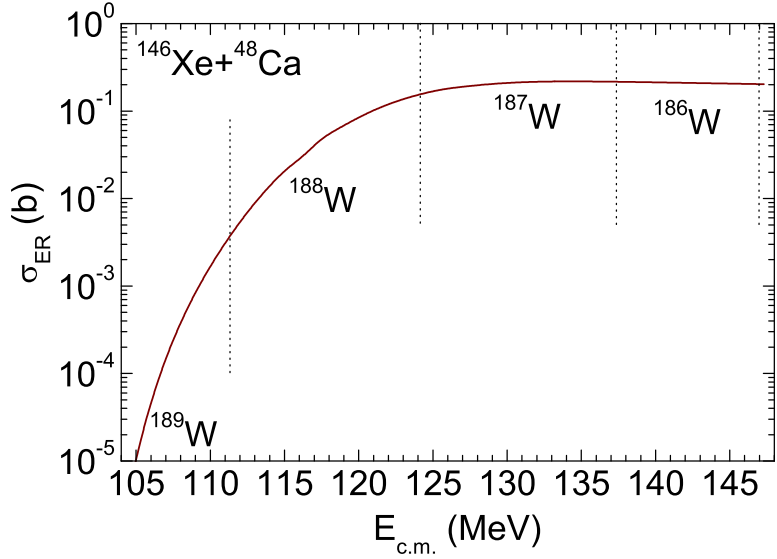}
\caption{The expected evaporation residue cross sections $\sigma_{ER}$ for
the indicated neutron-rich isotopes  $^{186-189}$W  produced
in the  $^{146}$Xe+$^{48}$Ca reaction.
The vertical dashed lines show the range of energies for the
production of  given isotope.}
\label{3_fig}
\end{figure}
\begin{figure}[tbp]
\includegraphics[scale=1]{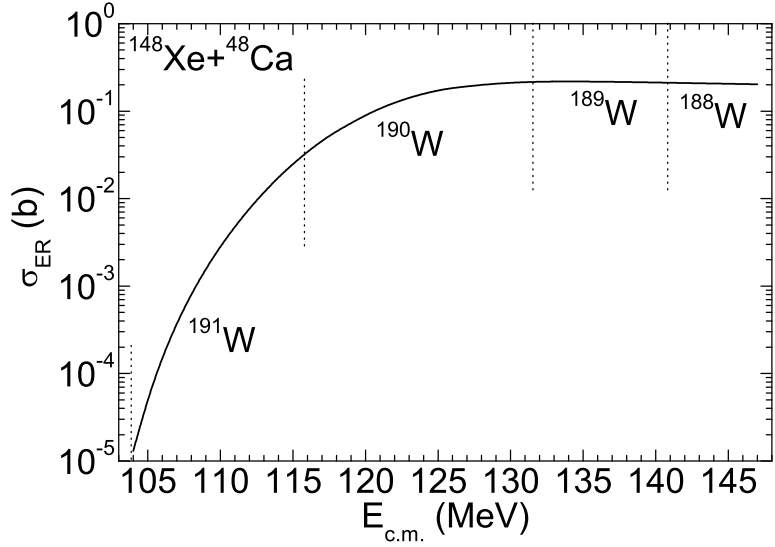}
\caption{The expected  evaporation residue  cross sections $\sigma_{ER}$ for
the indicated neutron-rich isotopes  $^{188-191}$W  produced
in the  $^{148}$Xe+$^{48}$Ca reaction.
The vertical dashed lines show the range of energies for the
production of  given isotope.}
\label{4_fig}
\end{figure}
In Fig.~1, one can see the comparison of the calculated functions $F(x)$
for the reactions
$^{130,132,134,136,138,140,142,144}$Xe+$^{48}$Ca
with stable and radioactive beams. As  expected, at sub-barrier energies
 the enhancement of the complete fusion (capture) cross section is larger in
the case of reactions with strongly quadrupole deformed projectile-nuclei and after neutron transfer.
The quadrupole deformation parameter $\beta_2$
of the projectile nucleus increases with changing mass number
from $A$=136 to $A$=130 or to $A$=144.
For the reaction   $^{136}$Xe+$^{48}$Ca  with spherical
target and projectile  and without neutron transfer the  cross section
is the smallest one at $x<0$. The sub-barrier cross sections for
the reactions $^{138,140,142,144,146,148,150}$Xe+$^{48}$Ca with neutron transfer (positive $Q$-values)
are larger than those for the reactions $^{130,132,134,136}$Xe+$^{48}$Ca,
where the neutron transfer is suppressed (negative $Q$-values).
Since after two-neutron  transfer the mass numbers and the deformation parameters
of the interacting nuclei are changed and the height
of the Coulomb barrier decreases,
one can expect an enhancement
of the capture.
For example, after the neutron
transfer in the reaction
$^{144}$Xe($\beta_2=0.18$)+$^{48}$Ca($\beta_2=0$)$\to ^{142}$Xe($\beta_2=0.15$)+$^{50}$Ca($\beta_2=0.25$),
the deformation of the target-nucleus increases and the
mass asymmetry of the system decreases,  and,
thus, the value of the Coulomb barrier decreases and
the capture cross section becomes larger (Fig.~1).
We observe the same behavior in the reactions with the projectiles $^{138,140,142,146,148,150}$Xe.

The complete fusion (capture) cross sections
for the reactions
$^{130,132,134,136,138,140,142,144,146,148,150}$Xe+$^{48}$Ca
at different bombarding energies
are presented in Fig.~2.
The behaviour of the curves in Fig.~2 is determined by
the quadrupole deformation and neutron transfer effects.
The isotopic dependency is rather weak at energies above the corresponding Coulomb barriers.
At sub-barrier energies
the  fusion cross section
decreases by
about one order of magnitude
with increasing mass number $A$ of the projectile from $A=130$ up to $A=136$ ($N=82$).
For $A > 136$ a steep increase can be observed for beam energies of
5  MeV below the corresponding Coulomb barriers.
At energies near the Coulomb barrier the cross section changes in a similar way but
the curve shows a much flatter slope.

\begin{figure}[tbp]
\includegraphics[scale=1]{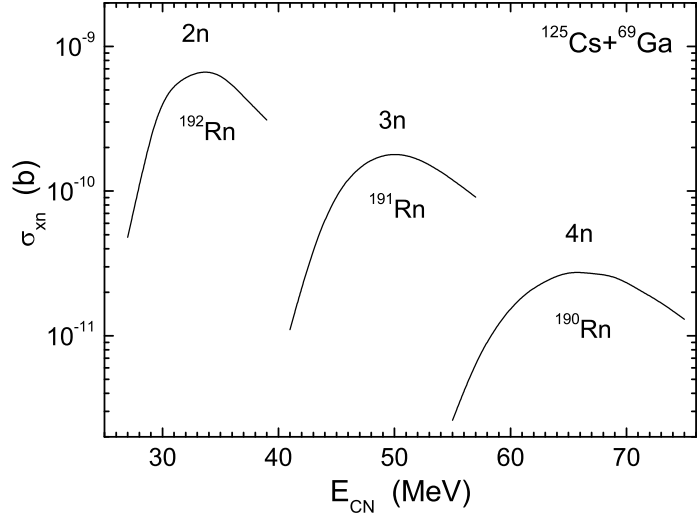}
\caption{The expected evaporation residue cross sections $\sigma_{xn}$ for
the indicated neutron-deficient isotopes of Rn  produced
in the  $xn$-channels ($x$=2-4) of the  $^{125}$Cs+$^{69}$Ga  reaction.
}
\label{5_fig}
\end{figure}
\begin{figure}[tbp]
\includegraphics[scale=1]{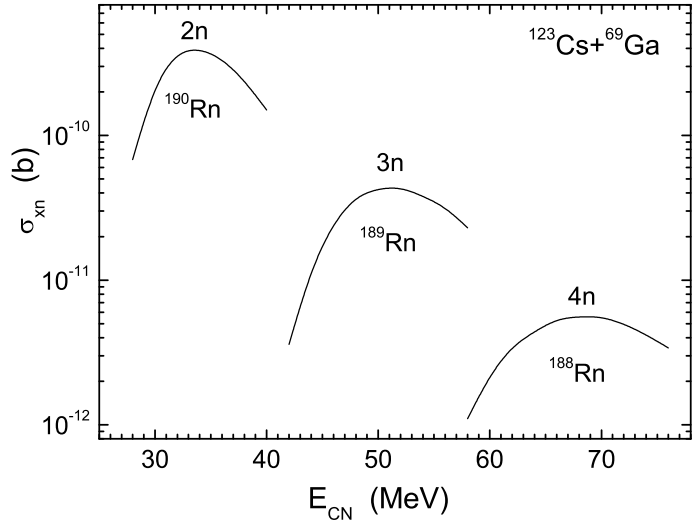}
\caption{The same as in Fig.~5, but for the  $^{123}$Cs+$^{69}$Ga  reaction.
}
\label{6_fig}
\end{figure}
In  Figs.~3 and 4
we present the possibilities for future experiments
to produce the neutron-rich isotopes $^{186-191}$W
in complete fusion
reactions  of $^{146,148}$Xe+$^{48}$Ca with radioactive beams.
The production cross sections of the neutron-rich  $^{190,191}$W isotopes, for example,
are between the 10 $\mu$b and 100 mb levels meaning that
they can be observed with rather low beam intensities and with the present experimental techniques.
The calculated cross sections are more than two orders of magnitude larger than in
fragmentation reactions \cite{Benlliure}.
Note also, that when the
neutron number approaches the drip-line
the production cross section in complete fusion decreases not so fast as in fragmentation reactions.

\subsection{Comparison between complete fusion and fragmentation reactions}
The availability of heavy radioactive beams at Coulomb barrier energies at future facilities like
FAIR, HIE-ISOLDE or SPIRAL-II will enable the experimental utilization of the above discussed effects
for fusion reactions.
Another competing method to produce heavy exotic isotopes is projectile fragmentation at
relativistic energies which is for example used at the Fragment Separator (FRS) at GSI.
In the following, we  give some comparative considerations on both methods since,
depending on the region of the nuclear chart, fragmentation can lead to high yields of exotic
nuclei.
As an example, we consider  the isotope $^{189}$W. Cross-sections of up
to about 2 mb are predicted for its production in the complete fusion reactions of
$^{146}$Xe+$^{48}$Ca at E$_{cm} = 110$ MeV.
Cross-sections on the same order are also measured in the fragmentation reactions
leading to yields of $10^4$ ions/s. At the future Super-FRS facility even yields of $2\times 10^6$ ions/s
are predicted. In order to obtain at least the same yields of $10^4$ ions/s in fusion reactions,
$^{146}$Xe beams
with intensities of at least $10^{13}$ ions/s are required. The largest intensities for neutron-rich Xe
beams are expected at SPIRAL-II where $10^5$ of $^{146}$Xe projectiles per second
are predicted which is, however, still eight orders of
magnitude less than needed for an efficient application of fusion reactions to reach $^{189}$W.

As an other example, we discuss in the following  the synthesis of neutron deficient Rn ($Z=86$) isotopes
in the complete fusion reactions.
Figures 5 and 6 show the calculated excitation functions for fusion reactions of $^{123,125}$Cs beams with
$^{69}$Ga target. The survival probabilities of the excited compound nuclei
in the neutron evaporation channels $xn$ ($x=2-4$) are calculated
 by employing the modified statistical
code GROGIF \cite{GROGIF} with the same parameters as in Ref. \cite{AZ}.
The capture cross sections and fusion probabilities are calculated with the quantum diffusion
approach \cite{EPJSub,EPJSub1}  and the dinuclear system fusion model \cite{AZ}, respectively.
Radioactive Cs beams are already now available with high intensities for a
broad variety of isotopes and are therefore favourable projectiles. At REX-ISOLDE for example,
the isotopes $^{122-129}$Cs are
provided with intensities around $10^{10}$ ions/s and for the future HIE-ISOLDE facility
ten times higher intensities are expected at beam energies of $\geq$ 5.5 MeV/nucleon.
A comparison of the predicted yields for neutron deficient Rn isotopes at the SuperFRS facility
with the expected yields from fusion evaporation reactions with $^{123}$Cs beams at intensities
of $10^{10}$ ions/s leads to the conclusion that the complete fusion is not superior to fragmentation for
$^{A}$Rn isotopes with $188\leq A\leq 190$. For these mass numbers at least 2-7 times lower yields can be
obtained in the fusion reactions  with the presently available beam intensities.

\section{Summary}
Because of deformation and neutron
transfer effects, a strong dependence of the sub-barrier complete fusion (capture) cross
section on the isospin was found for the reactions $^{130,132,134,136,138,140,142,144,146,148,150}$Xe+$^{48}$Ca.
At fixed bombarding energy, the cross section increases with changing mass number of the projectile-nucleus
from $A$=136 to $A$=130 or to $A$=150. The  $^{136}$Xe+$^{48}$Ca  reaction
with magic and semimagic nuclei has the smallest cross section.
The complete fusion (capture) cross sections for
the reactions $^{130,132,134,136}$Xe+$^{48}$Ca without
 neutron transfer are smaller than those for the reactions
$^{138,140,142,144,146,148,150}$Xe+$^{48}$Ca with  neutron transfer.
We demonstrated the possibilities for producing neutron-rich isotopes of
$^{186-191}$W with relatively large cross sections
 for future experiments in the complete fusion
reactions $^{146,148}$Xe+$^{48}$Ca with  radioactive beams. However,
we found that for the production of neutron-rich W
the fragmentation reactions are more preferable than the complete fusion reactions.
Even if we consider here the formation of   neutron-rich W isotopes as an example,
our findings have general validity and are not restricted to specific isotopes. Exotic nuclei with large
deformations which could be used as projectiles can equally be found in wide regions on the neutron-rich
as well as on the neutron-deficient side of the nuclear chart.

We  concluded also that the complete fusion  $^{123}$Cs+$^{69}$Ga reaction
with radioactive beam $^{123}$Cs is not superior to fragmentation for the production
of neutron-deficient isotopes of $^{188-190}$Rn.
The fragmentation reactions result in slightly larger yields of
these isotopes.
Note that the choice of the method of production of the isotopes near
the drip lines would be also affected by the purposes of the experiments
and the available facilities.

This work was supported in part by DFG and RFBR.
The IN2P3(France)-JINR(Dubna) and Polish - JINR(Dubna)
Cooperation Programmes are gratefully acknowledged.


\end{document}